%% file: main.tex
\documentclass[10pt,aps,prl,letterpaper,twocolumn,superscriptaddress,showpacs,amsmath,amssymb,nofootinbib,floatfix]{revtex4-1}
\usepackage{graphicx}  
\graphicspath{ {figures/} } 
\usepackage{bm}     
\usepackage{placeins}
\usepackage{amssymb}   
\usepackage{amsmath}   
\usepackage{physics}  
\usepackage[normalem]{ulem}   
\usepackage{xcolor}
\usepackage{mhchem}
\usepackage[utf8]{inputenc}
\usepackage[acronym,style=super,nonumberlist,toc]{glossaries} 
\usepackage{hyperref}
\usepackage{cleveref}
\usepackage{siunitx}
\usepackage{todonotes}

\begin{document}
\title{New Limits on \texorpdfstring{\ensuremath{n \rightarrow n'}}{n n'} Transformation from HFIR Cold Neutron Beam}

\author{James M. Rogers}
\affiliation{University of Tennessee, Knoxville, TN 37996, USA}
\author{Leah J. Broussard}
\affiliation{Oak Ridge National Laboratory, Oak Ridge, TN 37831, USA}
\author{Christopher B. Crawford}
\affiliation{University of Kentucky, Lexington, KY 40506, USA}
\author{Lisa DeBeer-Schmitt}
\affiliation{Oak Ridge National Laboratory, Oak Ridge, TN 37831, USA}
\author{Matthew J. Frost}
\affiliation{Oak Ridge National Laboratory, Oak Ridge, TN 37831, USA}
\author{Francisco M. Gonzalez}
\affiliation{Oak Ridge National Laboratory, Oak Ridge, TN 37831, USA}
\author{Carolyn O. Haviland}
\affiliation{University of Tennessee, Knoxville, TN 37996, USA}
\author{Lawrence Heilbronn}
\affiliation{University of Tennessee, Knoxville, TN 37996, USA}
\author{Erik B. Iverson}
\affiliation{Oak Ridge National Laboratory, Oak Ridge, TN 37831, USA}
\author{Yuri Kamyshkov}
\affiliation{University of Tennessee, Knoxville, TN 37996, USA}
\author{Mubasshir Khan}
\affiliation{University of Kentucky, Lexington, KY 40506, USA}
\author{Andrew Mullins}
\affiliation{University of Kentucky, Lexington, KY 40506, USA}
\author{David Milstead}
\affiliation{Department of Physics, Stockholm University, 106 91 Stockholm, Sweden}
\author{Linus B. Persson}
\affiliation{Department of Physics, Lund University, P.O. Box 118, 221 00 Lund, Sweden}
\author{Cary Rock}
\affiliation{University of Tennessee, Knoxville, TN 37996, USA}
\author{Valentina Santoro}
\affiliation{Department of Physics, Lund University, P.O. Box 118, 221 00 Lund, Sweden}
\affiliation{European Spallation Source ERIC, Partikelgatan 5, 224 84 Lund, Sweden}
\author{Alexander Saunders}
\affiliation{Oak Ridge National Laboratory, Oak Ridge, TN 37831, USA}
\author{Shaun Vavra}
\affiliation{University of Tennessee, Knoxville, TN 37996, USA}
\author{Nathan D. Whittington}
\affiliation{University of Tennessee, Knoxville, TN 37996, USA}
\date{July 1, 2026}
\begin{abstract}

Hypothetical neutron $n$ to sterile neutron $n'$ transformations would violate baryon number $\mathcal{B}$ and point to the nature of Dark Matter. 
We performed a new search for $n \rightarrow n'$ using an intense cold neutron beam from the High Flux Isotope Reactor at Oak Ridge National Laboratory. We used a theoretical model 
that describes the transformation $n \rightarrow n'$ with two parameters: a small mass difference $\Delta{m}$ between the interaction states $n$ and $n'$ and a mixing vacuum angle $\theta_0$. 
A thin absorbing cadmium wafer was used in the center of the superconducting 6.6~T magnet which provided a large gradient for the non-adiabatic $n \rightarrow n'$ transition. No signal was observed above background in the $^{3}\text{He}$ neutron detector 20 meters downstream of the magnet. This result gives an order of magnitude improvement in the lower limit for the probability $2\theta_0^2$ of the $n \rightarrow n'$ transformation in vacuum in the range of $\Delta{m}$ between $0.1$~neV and $1000$~neV. 
\footnote{This manuscript was authored in part by UT-Battelle, LLC, under the DE-AC05-00OR22725 contract with the US Department of Energy (DOE). The publisher acknowledges the US government license to provide public access under the DOE Public Access Plan (http://energy.gov/downloads/doe-public-access-plan).}
\footnote{Corresponding author Matthew Frost:frostmj@ornl.gov }

\end{abstract}
\maketitle

\input{Section_Text/introduction.tex}

\input{Section_Text/Beamline_and_Experiment.tex}

\input{Section_Text/beamintensity.tex}
\input{Section_Text/detectorcounts.tex}
\input{Section_Text/results.tex}
\input{Section_Text/conclusions.tex}
\input{Section_Text/acknowledgments}

\bibliography{references.bib}
\end{document}

%% file: Section_Text/introduction.tex
\section{Introduction}
\label{sec:intro}
The mysterious and elusive nature of dark matter remains the subject of many theoretical models and experiments, e.g. \cite{Boddy:2022knd, bozorgnia2024darkmattercandidatessearches}. The Mirror Matter model of dark matter \cite{Okun__2007,Berezhiani_2001,Mohapatra:2025bdl} posits that neutron can be a state of a mixture of an ordinary neutron and a mirror neutron $(n,n')^T$~\cite{Berezhiani_2006}. In the case where the masses of $n$ and $n'$ are not degenerate~\cite{Berezhiani_2019}, the limits have been placed in our previous experiments with neutron reappearance at the Spallation Neutron Source (SNS) \cite{Broussard:2021eyr,Gonzalez:2024dba}.

Theoretical overview and motivation for the search for the neutron-to-mirror-neutron transformation $n \rightarrow n'$ were discussed in detail in the paper ~\cite{Gonzalez:2024dba}. The two parameters of the theoretical model that describe the transformation of $n \rightarrow n'$ are the difference in the masses of the states $n'$ and $n$, i.e. $\Delta{m}=m_{n'}-m_{n}$ and the vacuum mixing angle $\theta_0$ of the states $n$ and $n'$. In vacuum, the system forms two eigenstates $\ket{n_1} = \cos \theta_0 \ket{n} + \sin \theta_0 \ket{n'}$ and $\ket{n_2} = -\sin \theta_0 \ket{n} + \cos \theta_0 \ket{n'}$ with two different energy eigenvalues. The parameter $\theta_0$ is related to the coupling constant $\epsilon_{nn'}$ in the Hamiltonian (see Eq. \ref{eqn:phen:simp_hamiltonian}) according to $\tan 2\theta_0 = 2\epsilon_{nn'}/{\Delta{m}}$. The characteristic oscillation time of the $n \rightarrow n'$ system in vacuum is then defined as $\tau_{nn'}=\hbar/\epsilon_{nn'}$. A description of the evolution of the system in the matter environment, including the presence of the magnetic field, the optical potential, and the absorption of the materials, can be found in ~\cite{Gonzalez:2024dba} and ~\cite{Kamyshkov:2021kzi}. The probability of the $n \rightarrow n'$ process can be calculated from the evolution of the density matrix of a two-level system with the following Hamiltonian.
 \begin{equation}
\label{eqn:phen:simp_hamiltonian}
    \mathcal{H} = \left( \begin{array}{cc}
        U - i W& \epsilon_{nn'} \\
        \epsilon_{nn'} & 0
    \end{array} \right)\,,
\end{equation}
where $U$ includes all real quantities:
\begin{equation}
\label{eqn:phen:Udef}
U= V_F- \Delta{m} + \mu_{n} \vec{\sigma} \cdot \vec{B},
\end{equation}
where $V_F$ is the optical potential of the medium, $\mu_n$ is the neutron magnetic moment, $\vec{B}$ is the local magnetic field, and $\vec{\sigma}$
are the Pauli matrices. The potentials $U$ and $W$ interact only with the ordinary neutron component $n$, but not with $n'$. For the latter, we assume that the mirror magnetic field $\vec{B}'$ and the mirror optical potentials are absent. The evolution of the two-state $(n,n')^T$ system follows the Schr\"{o}dinger equation as in~\cite{Kamyshkov:2021kzi}. Any real $U > \epsilon_{nn'}$ leads to suppression of the $n \leftrightarrow n'$ transformation. $W$ is responsible for the absorption of the $(n,n')^T$ system in the material. The elastic scattering of neutrons in the beam transmission experiment ``measures" the $(n,n')^T$ system as a neutron that is lost from the beam. 

The limits on the probability of transformation $n \rightarrow n'$ in terms of the regions of the excluded parameters of the model $\Delta{m}$ and $\theta_0$ were obtained in our experiments~\cite{Broussard:2021eyr,Gonzalez:2024dba} carried out with the cold neutron beam at the Spallation Neutron Source (SNS) at Oak Ridge National Laboratory (ORNL). The first of these experiments has excluded the proposition made in~\cite{Berezhiani_2019} that this model of the $n \rightarrow n'$ transformation can explain the standing neutron lifetime controversy. The second experiment~\cite{Gonzalez:2024dba} significantly improved the limits of the model parameters. 

Here, we report on the new study performed with the CG-2 cold neutron beam from the High Flux Isotope Reactor (HFIR) at ORNL. For this study, we used the same type of superconducting magnet that we used in SNS experiments~\cite{Broussard:2021eyr,Gonzalez:2024dba}. The advantages of using the HFIR cold neutron beam compared to SNS were the following: 400 times higher average beam intensity, a wider and colder spectrum of neutrons, and the low-background $^{3}\text{He}$ tube-style neutron detector of the GP-SANS instrument~\cite{BERRY2012179}.

%% file: Section_Text/Beamline_and_Experiment.tex
\section{GP-SANS Beamline and Experiment} 
\label{sec:experiment}
The General-Purpose Small Angle Neutron Scattering Instrument (GP-SANS)
 is located at the CG-2 beam port within the Cold Guide Hall
of the HFIR reactor operated by ORNL. The high-intensity CG-2 beam
comes from a 20~K cold liquid hydrogen source installed
close to the core of the HFIR reactor~\cite{HFIRCS2018}. The neutrons
from the cold source are extracted through a 4~m long $4\times 4~$cm$^2$
nickel-coated collimator guide from the main shutter to the ``optical filter'', which is an $m = 3$ reflector,
changing the initial direction of the cold neutrons in the horizontal plane by $2^{\circ}$ and providing a fast neutron cutoff of $\lambda < 2.5~\text{\AA}$. 
 The neutron wavelength used in this measurement ranged from 2.5 to 20 $\text {\AA}$ and the average wavelength of neutrons in this range was $\bar{\lambda} \sim 5.42~\text{\AA}$.
After the optical filter, neutrons travel through two segmented Ni $m=1$ vacuum guides
with aperture $4\times 4~$cm$^2$ that are 19~m and 16~m long. 
At the end of the second guide, the root mean square (RMS) of the divergence of the cold beam was $0.19^{\circ}$ in the horizontal and $0.26^{\circ}$ in the vertical direction.
The details of the resulting beam spectra, attempts to account for any dependencies between modeled and observed performance, and an up to date layout of the incident optics are provided in ~\cite{Rogers:2024mds}.

\begin{figure}
    \centering
    \includegraphics[width=\columnwidth]{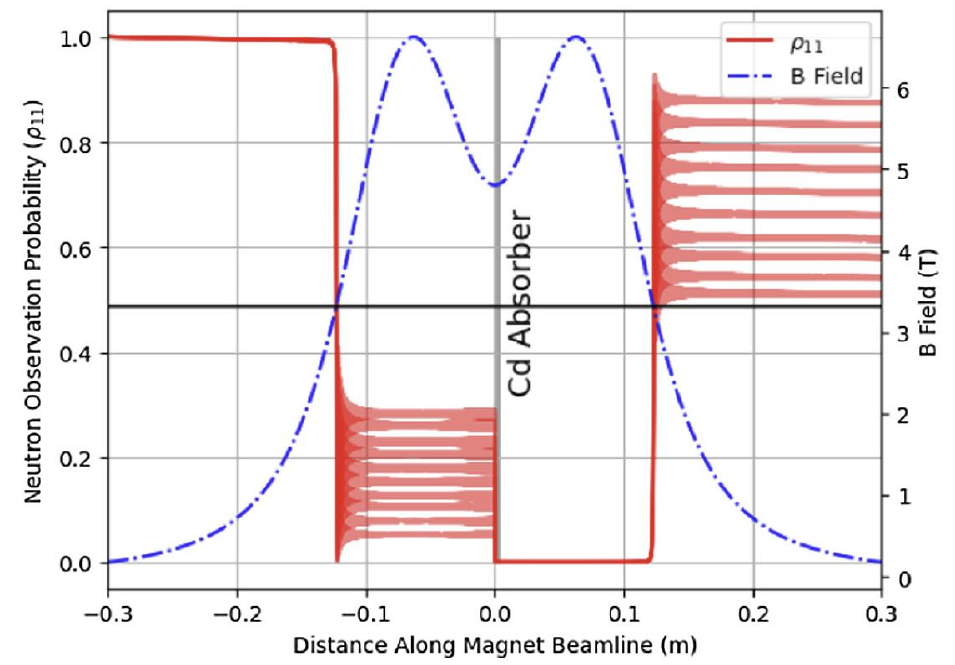}
    \caption{borrowed from paper~\cite{Gonzalez:2024dba}. Blue dot-dashed line shows the total magnetic field (right scale) along the 
    beam axis and the position of 3.5 mm thick cadmium absorber inside the magnet. Red 
    lines give an example of simulated evolution of probabilities (left scale) of 
    detectable neutrons for 10 different initial velocites due to 
    non-adiabatic Landau-Zener transition ~\cite{Landau:1932, Zener:1932ws} 
    for the theory parameters $\Delta{m} = 200$\,neV, and 
    $\theta_0 = 5\times10^{-3}$ excluded by our previous experiments~\cite{Broussard:2021eyr,Gonzalez:2024dba}. 
    The red line is the square of the neutron wave function, which gives the probability of neutron observation shown for each velocity sampled from the spectrum distribution (color online). The black horizontal line crosses the magnetic field that would 
    correspond to $\Delta{m}$ in the calculation. }
    \label{fig:magprofile}
\end{figure}

From the end of the last neutron guide, the vacuum continues for $\sim 1$~m, ending with a 6.35 mm thick sapphire window immediately followed by a 20 mm diameter boron nitride aperture in air.
Another 20 mm diameter boron nitride aperture is installed in front of the magnet at a distance of 155 mm from the first aperture.
Following this aperture, a thick-walled boron carbide cylinder (inner diameter 25~mm, length 150~mm) is installed in the first half of the magnet. A 3.5~mm cadmium wafer is mounted at its downstream end, at the center of the magnet, where it absorbs the cold neutron beam.
The superconducting split-pair magnet with maximum field 6.6~T, from 
Cryomagnetics, Inc.~\cite{Magnet}, is similar to the magnet used in our previous
$n \rightarrow n'$ search experiments at the SNS~\cite{Broussard:2021eyr,Gonzalez:2024dba}. 
The magnet is 30~cm in length along the beam axis with the cadmium absorber positioned in the center. 
The magnet is installed in the air gap of 57 cm between the sapphire window and a 1 cm thick
silicon window of the 20 m long vacuum tank of the GP-SANS $^3$He neutron detector ~\cite{BERRY2012179}.
The distance between the end of the neutron guide
and the center of the magnet is 1.3 m through the vacuum, the sapphire window, 
and the air. The profile of the total magnet field is shown as a blue
dot-dashed line in Figure~\ref{fig:magprofile}.

The characteristics of the neutron detector, including geometry, resolution, efficiency, etc., as a standard instrument of the GP-SANS user facility, are discussed in \cite{BERRY2012179}. 
The detector consists of two planes of 8-mm diameter thin stainless steel tubes of one meter length arranged vertically with 11~mm tube spacing.
The horizontal $x$ coordinate of the neutron interaction with $^{3}$He is reconstructed from the number of tubes and the vertical coordinate $y$ along the length of the tubes by the division of charges between two signals measured at both ends of the resistive wire of the tube. 
The vertical resolution of the detector is 7 mm. The tubes operate in proportional mode. 

The neutron beam is unpolarized.
For one of the neutron spin states, the presence of a strong magnetic field allows compensating suppression of the hypothetical $\Delta{m}$ difference between the $n$ and $n'$ states in the first half of the superconducting magnet and leads to the enhancement of the $n \rightarrow n'$ transformation. The 3.5 mm thick cadmium wafer installed in the middle of the magnet completely absorbs all beam neutrons (to the level below $10^{-20}$) but allowed the $n'$s to travel through cadmium without interactions and then be transformed back into detectable neutrons in the field of the second half of the magnet. This method of ``beam regeneration'' leverages the advantages of a low-background neutron detector for the detection of regenerated neutrons.

\begin{figure}
     \centering
     \includegraphics[width=\columnwidth]{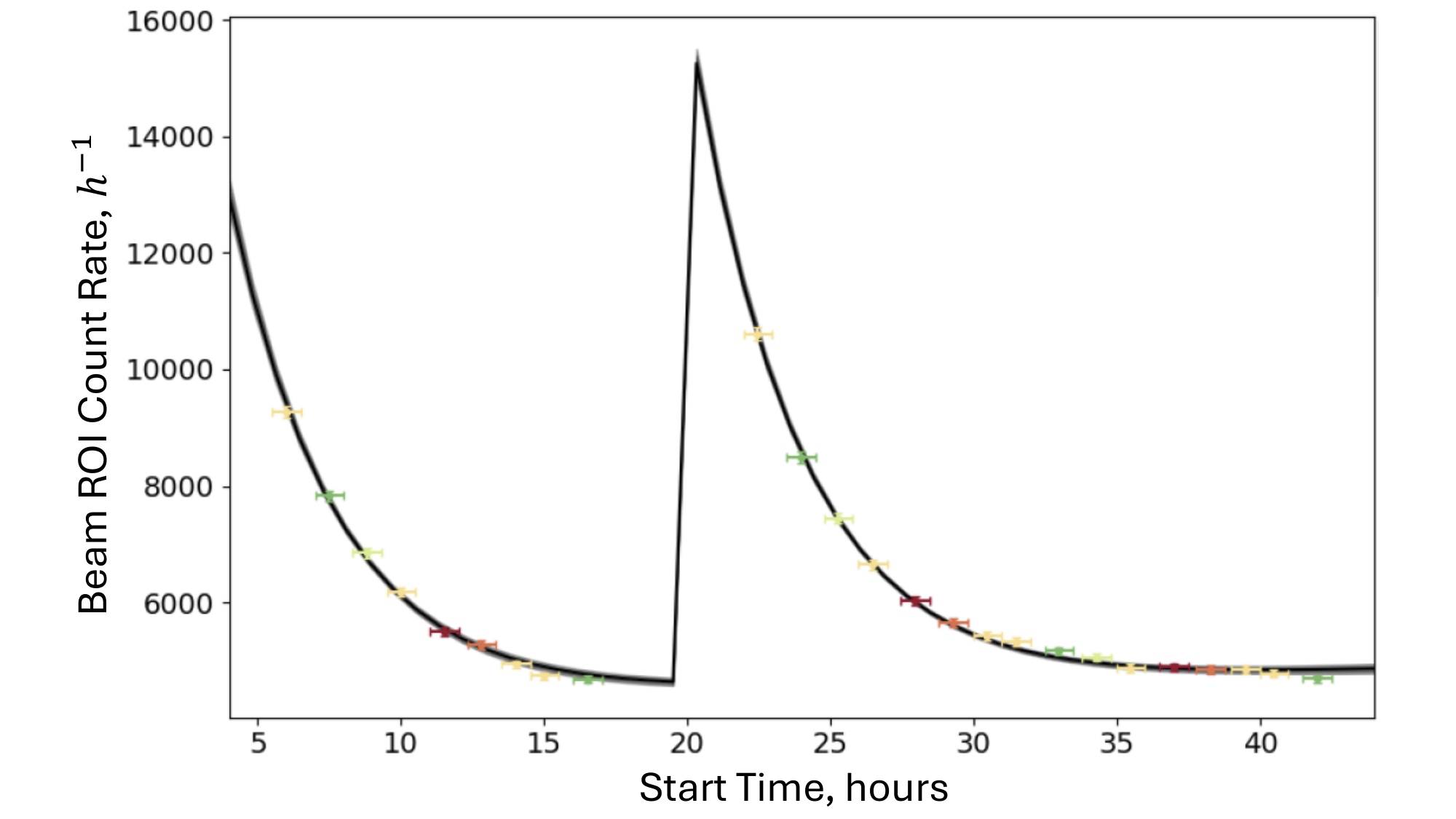}
     \caption{Background count rate in ROI (black line) and measurement periods (alternating colored bars) during the 2021 measurement. The detector was activated by the neutron beam at the start of the run and at around hour 20. The exponential change of detector counting rate was attributed to the decay of activated $^{56}$Mn, the activation product of $^{55}$Mn present at the level 
     of $\sim$2\% in the stainless steel material in tubes of the detector.}
     \label{fig:background2021}
 \end{figure}

Measurements were performed for $\sim$40 hours of beam time in 2021 and for $\sim$90 hours in 2024. In 2021, as part of a beam intensity calibration measurement, we briefly allowed a relatively high neutron intensity in the detector. We found that the detector was activated, increasing the background rate, with an exponential decay time constant of $\sim$3.7 hours after irradiation (Figure \ref{fig:background2021}).
The increase of background was assumed to be related to the activation of $^{55}$Mn, which makes up $\sim$2\% of stainless steel walls in the $^{3}$He detector tubes.
In addition, the detector threshold was not high enough
to suppress the detection of $\gamma$ rays produced by decaying $^{56}$Mn. These factors prevented the GP-SANS neutron detector from reaching its lowest background level during this measurement period. However, the 2021 data were analyzed and weighted in the overall result.

During the 2024 run, we were careful not to allow high beam intensity in the detector
except for a short 30~s exposure with attenuated intensity at the end of the run for measurements of the beam footprint in the detector. The typical background in the \SI{1}{m^2} detector in the 2024 run with the cadmium absorber inside the magnet was $\sim$ 15~cps. The count rate in the region of interest (ROI) with size $\sim$$20 \times 20$ cm$^2$ was typically 1 cps. Due to a higher background, the ROI  in 2021 was optimized to a smaller size, where the variable background was 1.5--3.5~cps. 

The measurements were arranged in one-hour detector counting periods with different currents in the superconducting magnet. These periods were alternated with periods of magnetic fields ramping up and down, and of $\sim$50\% of one-hour periods with nominally zero magnetic field \footnote{The magnetic field in these periods was actually $\sim$\SI{0.003}{T}.} for background measurement. The beam intensity was continuously monitored by the calibrated low-efficiency nitrogen-filled counter ORDELA 4511N~\cite{ORDELA}, with efficiency $9.73 \times 10^{-6}$ at $\lambda = 1.8~\text{\AA}$, installed in the air gap between the 19-m and 16-m neutron guides. A similar ORDELA counter with efficiency ${\sim}10^{-5}$ was also used for a few short measurements directly in front and behind the superconducting magnet to account for the effect of beam intensity reduction due to trimming the beam size from $4 \times 4~$cm$^2$ to the aperture with a diameter of 2 cm in front of the magnet. In 2021, two magnetic field settings were used: 100\% and 50\% of the maximum field; in 2024 the settings were 100\%, 75\%, and 50\% of the reachable maximum. The information from the GP-SANS neutron detector and the monitors for every neutron interaction was recorded by EPICS-based DAQ software~\cite{EPICS}.

%% file: Section_Text/beamintensity.tex
\section{Beam Intensity Measurement} 
\label{sec:intensity}
For measurements of the absolute beam intensity of the cold neutron beam with spectrum from 2.5 to 20~\AA~\cite{Rogers:2024mds}, we used beam equipment already existing within the HFIR research facility. To look for the effect of systematic uncertainties, e.g., in the quoted monitor detection efficiency, we performed measurements with two different devices or methods in 2021 and 2024.

In the 2024 run, the intensity of the cold neutron beam propagating through the neutron guide with an aperture of $4 \times 4~\text{cm}^2$ was measured and monitored with ORDELA~\cite{ORDELA} N$_2$-gas counters with calibrated efficiency $9.73\times 10^{-6}$ at $\lambda = \SI{1.8}{\angstrom}$. 
The typical count rate of this detector, called the GP-SANS monitor or GPM, was ${\sim}1.107 \times10^5~\text{s}^{-1}$. 
The electronics of the ORDELA detectors had a paralyzable dead time and was also prone to double pulsing, sensitive to the threshold.
Therefore, to correct for significant dead-time effects, we used the distribution of time intervals between discriminated signals. 

\begin{figure}
     \centering
     \includegraphics[width=\columnwidth]{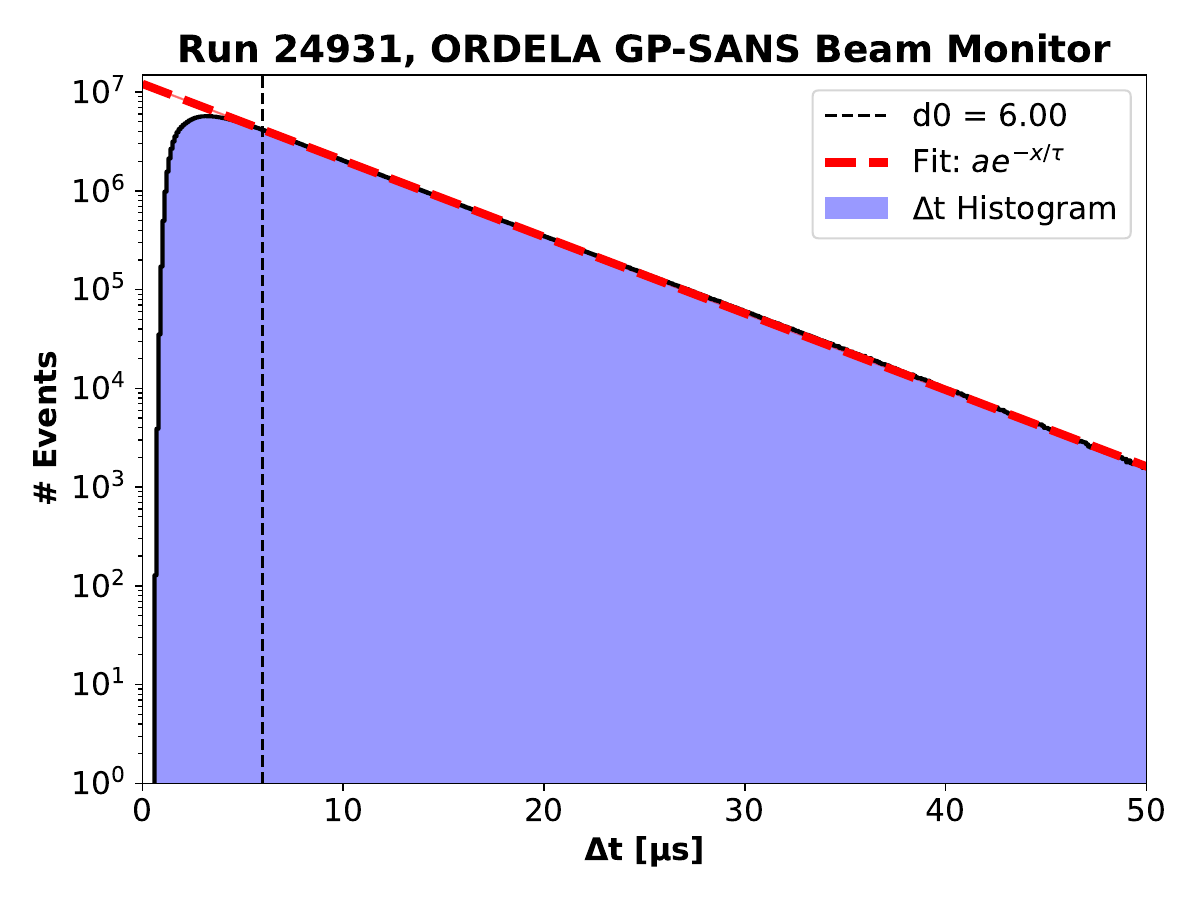}
     \caption{Distribution of time intervals between signals from ORDELA GP-SANS beam monitor for one-hour measurement in 2021. An exponential slope is a fit to the histogram data for which $\Delta t > d_0 = 6.00 \mu s$, set to exclude the dead time region of small $\Delta t$ and afterpulsing.}
     \label{fig:time_intervals}
 \end{figure}

The EPICS system recorded the GPS-synchronized time of each signal with precision $< \SI{100}{ns}$. 
For events of random Poisson origin, the distribution of time intervals follows purely exponential dependence. 
An example of such a distribution is shown in Figure \ref{fig:time_intervals}. 
The slope of the distribution is related to the average count rate, with distortion evident below $\SI{6}{\micro\second}$ due to deadtime and afterpulsing. To correct for deadtime, the distribution was fit to an exponential above $6.00 \mu s$, and the missing counts below $6.00 \mu s$ was calculated from the fit.
The correction factor due to the dead time of the GP-SANS monitor in 2024 was typically $1.146 \pm 0.011$ and varied within ${\sim}1\%$ from run to run. The precision of the correction factor determined from the exponential fit to a single run was $0.07\%$. The detection efficiency of the monitor was proportional to the neutron wavelength of the spectrum, with $\bar{\lambda}=5.42 \text{\AA}$ averaged over the neutron spectrum. Thus, the average beam intensity measured by the GP-SANS monitor corrected for deadtime and detection efficiency was $(4.08 \pm 0.04) \times 10^9$ n/s, varying within $4\%$ over the runs in the 2024 measurement. 
 
\begin{figure}
     \centering
     \includegraphics[width=\columnwidth]{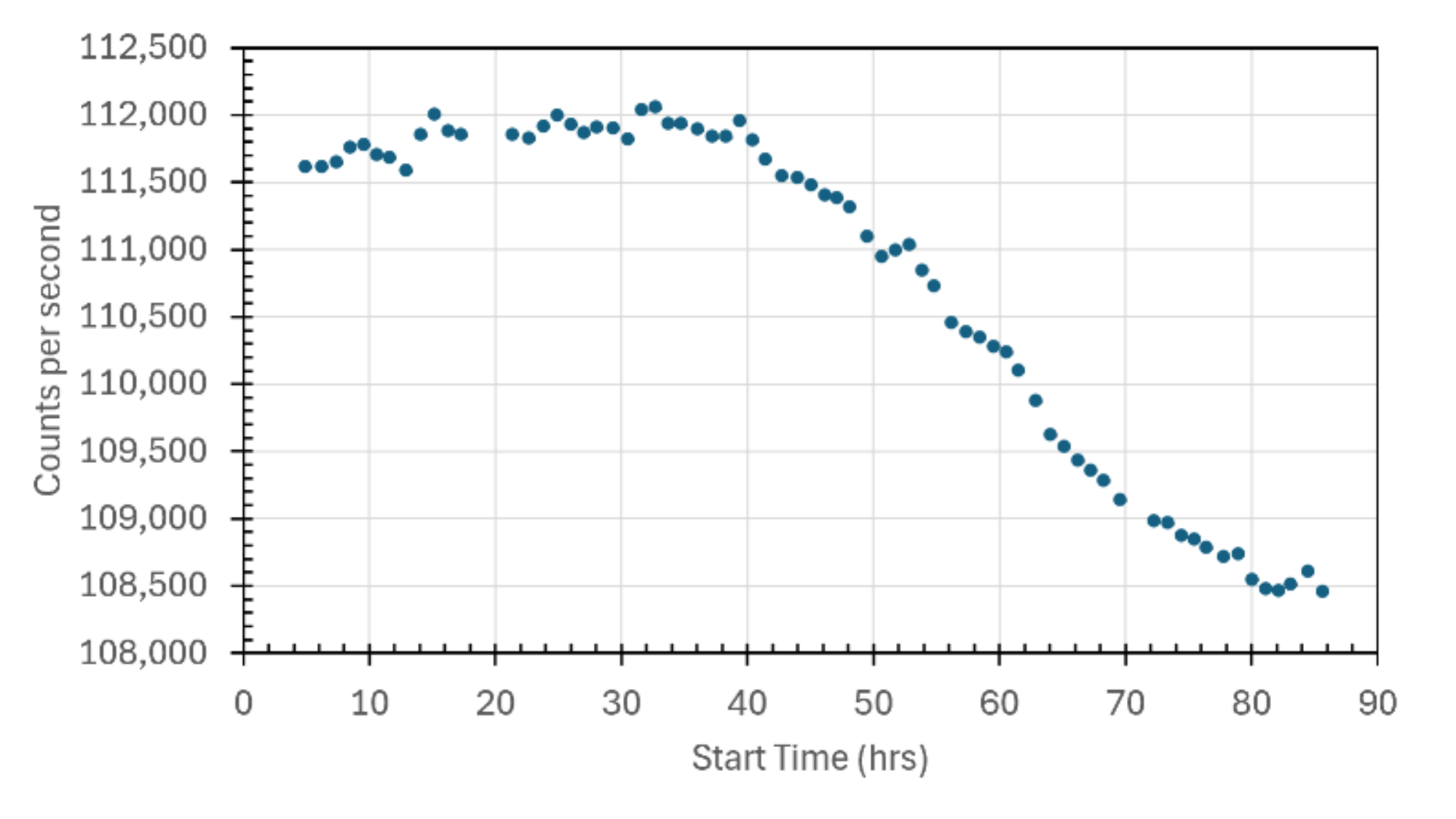}
     \caption{Average counted intensity of GP-SANS ORDELA monitor in all 1-hour runs in 2024. The statistical errors of monitoring measurement are withing the point size on the figure. Steady variation of monitored beam intensity for 90-hours beam time is 3.2\%.}
     \label{fig:intensity}
 \end{figure}

The count rate of the GP-SANS monitor slowly changed during data collection because of the variation of the reactor intensity. The variation for the 90 hour run in 2024 is shown in Figure~\ref{fig:intensity}. Here, the beam intensity gradually changed run-by-run  within $\sim$3.2\% for 90 hours. 
This intensity variation was used for the correction for the normalization of the main detector counts. 

A circular aperture of 2-cm diameter was placed in front of the magnet, which reduced the beam cross section from a nominally square 
$4 \times 4$ cm$^2$ cross section to a nominally circular beam area of $\pi~\text{cm}^2$. The actual reduction in intensity due to these apertures was calculated with the McStas neutron ray-trace simulation package \cite{NISTA&S}, which determined an intensity reduction factor of 0.187 $\pm$ 0.011 with uncertainty including guide alignment systematics.

The effects of neutron scattering and absorption in two sapphire vacuum windows downstream of the GP-SANS monitor were calculated with data from the NIST calculator \cite{NISTA&S}; together with coherent scattering in the crystal with the C-axis orientation estimated by McStas, the transmission was $0.81 \pm 0.05$.
Combining these factors, the observed average beam intensity at the entrance of the magnet was $(6.18 \pm 0.47) \times 10^8$ n/s. The overall measured intensity variation mentioned above throughout the duration of the experiment was 4\% due to the intensity variation of the reactor and the uncertainty of dead time correction determination.

The beam intensity in 2024 was also measured directly by another ORDELA low-efficiency monitor alternatively positioned before and after the magnet.  Although count rates in these measurements were factor of $\sim$5 lower than in GPM, the dead time correction was $1.105 \pm 0.004$, due to the different paralyzing and afterpulsing properties of the counter. The average intensity measured by this counter, corrected for dead time and detector efficiency, was $(6.02 \pm 0.09) \times 10^8$ neutrons per second, consistent with the intensity reconstructed from the GP-SANS monitor. The unweighted average of the two intensity measurement methods results in the final intensity of $(6.10 \pm 0.24) \times 10^8$ or a precision of $4\%$ during the 2024 run.

In 2021, the beam intensity was defined using the GP-SANS monitor (GPM) with similar calculated reduction factors. This monitor was a different ORDELA unit than in 2024, with different dead time, threshold, and afterpulsing. The intensity measured by this monitor, with dead time and efficiency corrections, was $(6.03 \pm 0.05) \times 10^9$ n/s, slowly decreasing by $1.5 \%$ during the 40-hour period. After the transport corrections, the useful intensity at the entrance of the magnet was $(9.13 \pm 0.70) \times 10^8$ neutrons per second. 

An additional intensity measurement was performed in 2021 using gold foil activation. The foil was placed at the exit of the magnet with the cadmium absorber removed. The foil area did not cover the whole beam aperture and the uncertainty of coverage was 
${\sim}10\%$. This measured intensity was $(7.403 \pm0.756) \times 10^8$ n/s, a difference of about 1.7 standard deviations from the intensity measured by the monitor. By performing a weighted average of the two methods, the beam intensity at the magnet entrance in 2021 was found to be $(8.27 \pm 0.52) \times 10^8$ n/s, with an error of $6.3\%$. We noticed that the beam intensity in 2021 was a factor of $1.36 \pm 0.10$  higher than in 2024. 

%% file: Section_Text/detectorcounts.tex
\section{Detector counts and results} \label{sec:detector}
\begin{figure}
     \centering
     \includegraphics[width=\columnwidth]{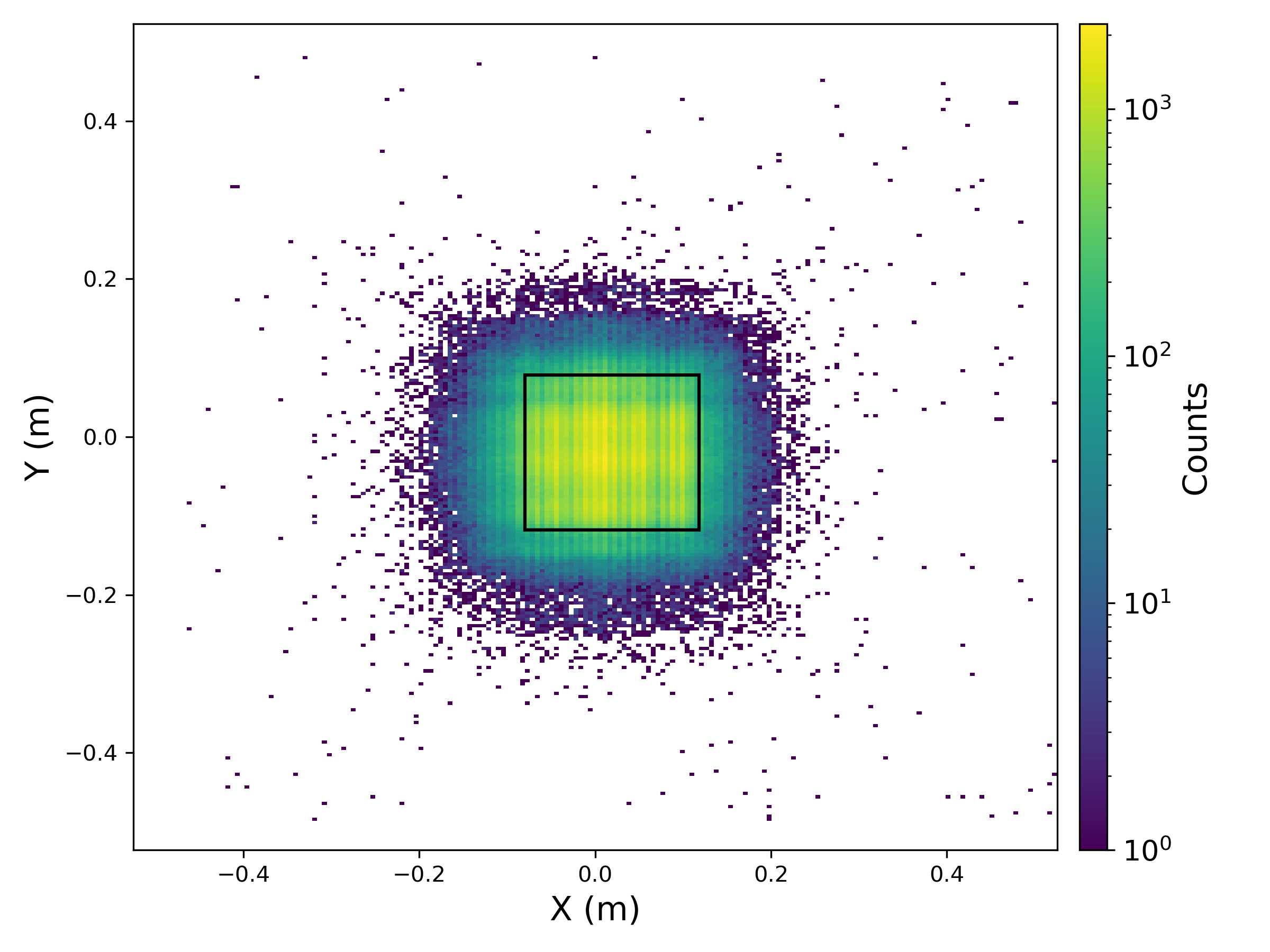}
     \caption{Footprint of the divergent beam on the GP-SANS detector in 2024. Black framed square is optimized region of interest (ROI) with size $20 \times 20~cm^2$ used for effect counting with efficiency 87.9\%.}
     \label{fig:footprint}
 \end{figure}

The GP-SANS detector was positioned at a distance of 20~m from the center of the magnet where the cadmium absorber was installed. In 2024, the beam profile in the detector was measured at the end of the run in a short 30-second exposure to the attenuated beam with the cadmium absorber removed. The footprint of the divergent beam on the detector is shown in Figure \ref{fig:footprint}. The $20 \times 20$~cm$^2$  region of interest (ROI) used to count regenerated neutrons is shown as the black framed square. The ROI size was selected by optimizing the signal-to-background ratio using footprint of the beam measured in dedicated short run without the cadmium absorber. The optimal ROI cutoff efficiency was determined to be 87.9\%. In 2021, the optimized ROI was smaller due to the higher background and had an efficiency of 77.7\%.

\begin{figure}
     \centering
     \includegraphics[width=\columnwidth]{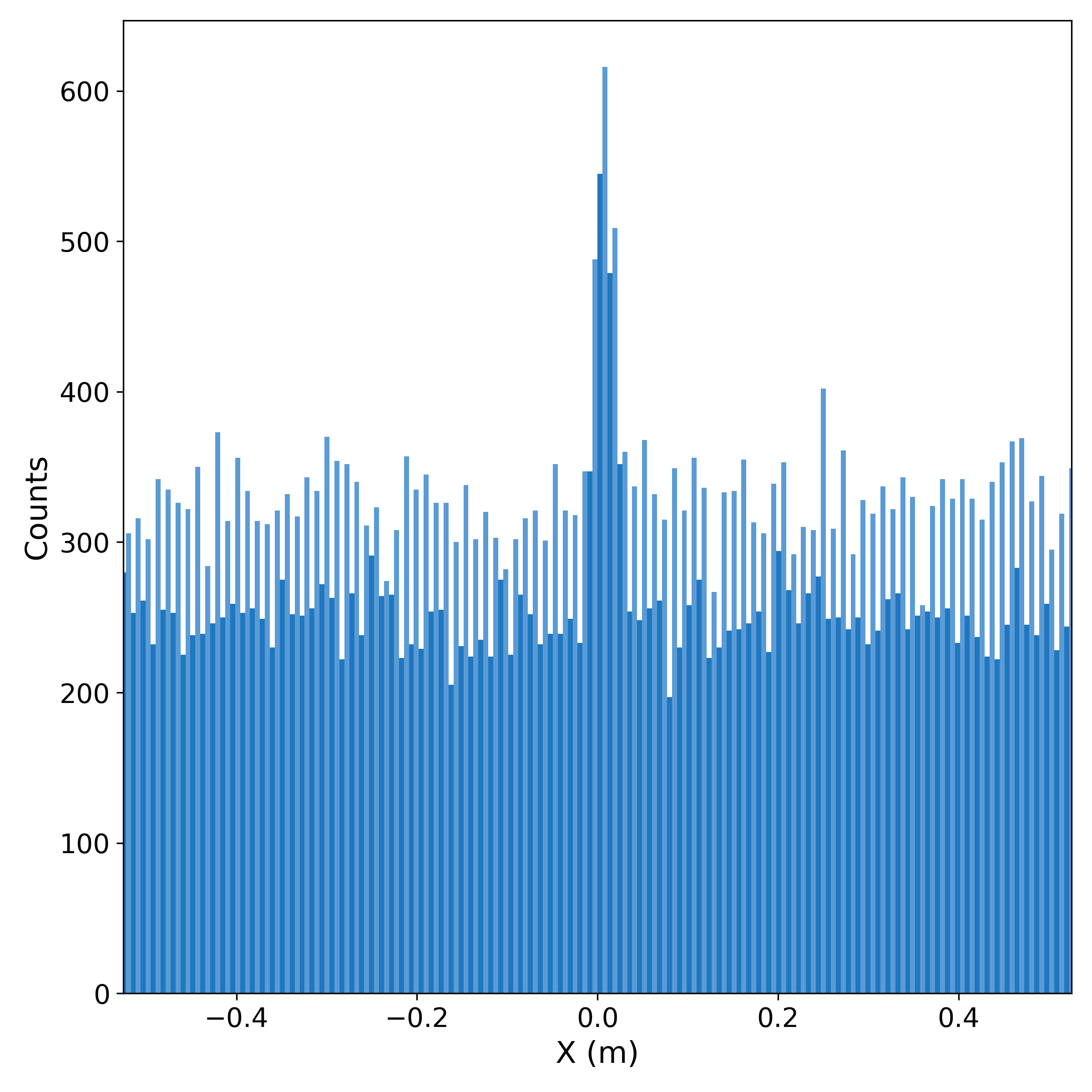}
     \caption{Distribution of background hits in the tubes of two layers of GP-SANS detector for a typical one-hour $B=0$ measurement with cadmium absorber in the beam. Front-row tubes are shown in light blue and back-row tubes in dark blue. Peak in the center is due to fast neutrons scattered from the $m=3$ ``optical filter'' reflector and coming through the straight 35-m neutron guide.}
     \label{fig:background}
 \end{figure}

The background in the detector for a typical one-hour counting period in the 2024 run with the magnetic field $B=0$ and the cadmium absorber in place is shown in Figure \ref{fig:background}. The central narrow peak in the background distribution is due to scattered high-energy neutrons from the reflection filter upstream of the long neutron guide punching through the cadmium absorber. This peak was present in all measurements with all values of the magnetic field and was treated as part of the background. 

The average count rate of the entire 1~m$^2$ detector in 2024 was $\sim$ 15 cps. The count rate within ROI was $\sim$1 cps. The detector background count rate measured in the ROI was increasing almost linearly with time at ${\sim}8\%$ during a 90-hour run, although the monitored beam intensity has been reduced by ${\sim}3.2\%$.
We attribute that the increase to the accumulation of the neutron activation background in the detector and in the vacuum tank materials.

Since the background was time dependent and measurements of the background and the effect for the different values of magnetic field were performed at different times, the following procedure for background subtraction was used. All background measurements made with zero magnetic field $B=0$ and with statistics of counts within the ROI were fitted with a linear-time dependence. The covariance matrix of this fit was used to generate $10^6$ random samples of the background in the ROI at times corresponding to measurements with parameters randomly taken from a multivariate normal distribution \cite{multinormal} with the given fit covariance matrix. The distribution of random sampled backgrounds was used to find the expected value for the background with its uncertainty at that measurement time. The expected value for the background with its error was then 
subtracted from the counts measured in ROI to determine the ``effect". 

\begin{figure}
     \centering
     \includegraphics[width=\columnwidth]{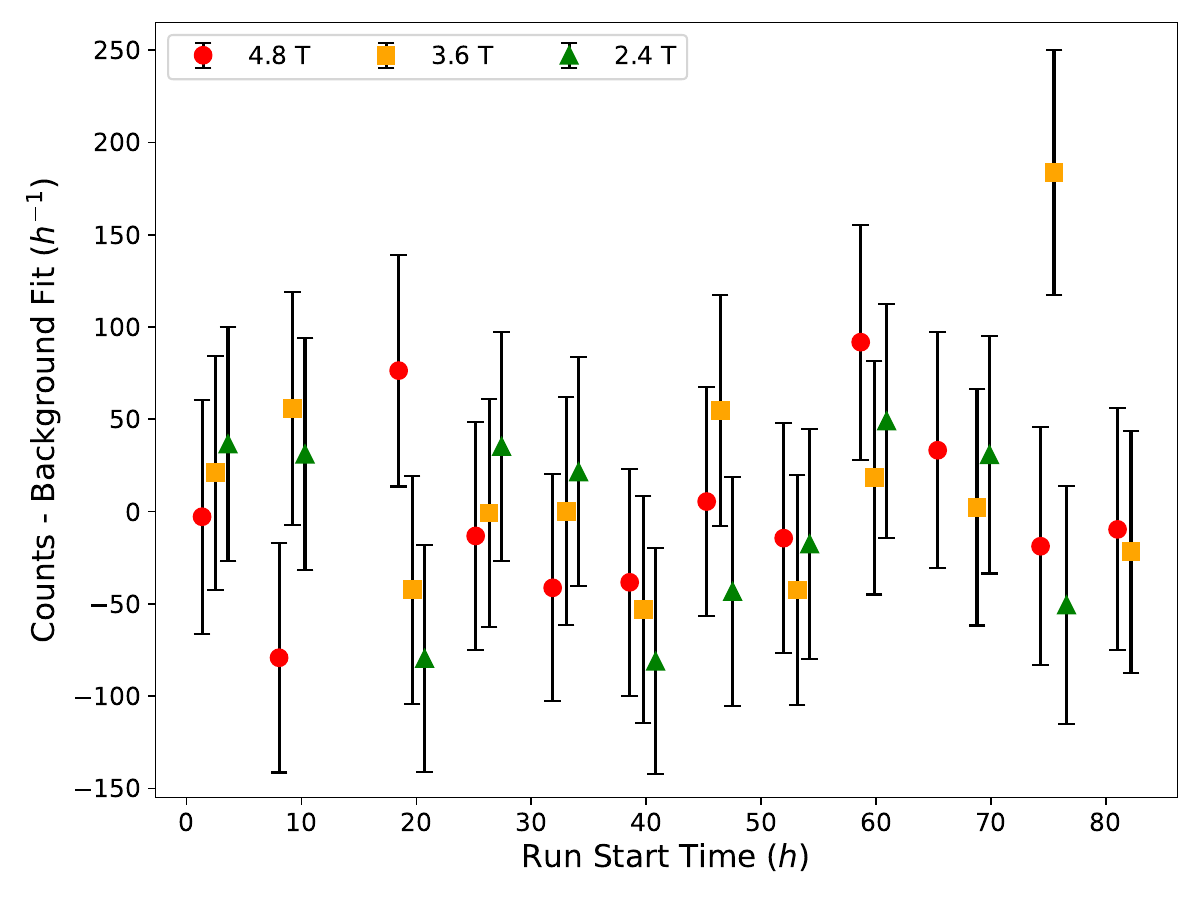}
     \caption{Regeneration effect for measurements with different values of magnetic field in 2024 data. The label of magnetic field is referred to the value in the magnet center. Background measured at $B=0$ was subtracted as explained in the text (color online).}
     \label{fig:effect2024}
 \end{figure}

The result of these subtractions for the 2024 data is shown in Figure \ref{fig:effect2024}. The points are consistent for measurements with positive and negative values of magnetic field (six one-hour measurements at each value of magnetic field) and globally consistent with a zero-effect observation. 
We averaged the counts of multiple measurements for each of three values of the magnetic field and calculated the corresponding probabilities and the limits of the effect at the 95\% CL using the Feldman Cousins method~\cite{Feldman:1997qc}. These limits are shown in Table \ref{tab:counts2024}. 

In the calculation of the neutron intensity at the GP-SANS detector, the estimated \cite{NISTA&S} $91 \pm 2$\% efficiency of transmission of a 1~cm thick silicon window of the neutron detector tank, the $96 \pm 1\%$ efficiency of the neutron detector averaged over the beam spectrum, and the $87.9 \pm 0.5\%$ efficiency of the ROI selection cut were taken into account. The general normalization factor was $(1.43 \pm 0.07) \times 10^{12} ~$h$^{-1}$, representing the total number of neutrons that could have hit the detector if 100\% were regenerated. The detected counts were divided by this factor to give the size of the signal 
in Table \ref{tab:counts2024} for conversion to the probability of effect per average neutron in the beam with definite polarization.

Since $n \leftrightarrow n'$ oscillations could also occur in the absence of magnetic field, we additionally treated the 33 one-hour measurements with $B=0$ in a way allowing us to extract the oscillation limit in vacuum. For that, the large area of the neutron detector outside the ROI with apparently uniform background density was used in each measurement to determine the background rate per unit area, and the corresponding counts were subtracted from the ROI. In this analysis the area $10 \times 10~cm^2$ around the fast-neutron peak shown in Figure \ref{fig:background} was removed from the definition of ROI. After subtracting, the average of 33 measurements was 24.4 counts per hour with a conservative error estimate of $\pm$ 49.2 taken from the root mean squared error. The intensity normalization factor for this result included the reduced efficiency of ROI down to $52.4\%$ due to the removal of the central peak and the fact that both polarizations of neutrons in the beam will produce an equal polarization effect in the absence of magnetic field. This normalization factor was $(4.79 \pm 0.23) \times 10^{12} ~$h$^{-1}$. The effects per neutron and the corresponding Feldman-Cousins 95\% CL limis for all values of magnetic field are shown in Table \ref{tab:counts2024}. 

\begin{table}[ht]
    \centering
    \begin{tabular}{|c|c|c|c|}
        \hline
        B-Field (T) & Counts, $h^{-1}$ & Signal ($\times 10^{-12}$) & $95$\% CL\\
        \hline
        $ 0.0 $ & $24.4 \pm 49.2$ & $5.1 \pm 10.3$ & $1.25 \times 10^{-11}$\\
        $ 2.40$ & $-9 \pm 21$ & $-6.34 \pm 14.7$ & $2.28 \times 10^{-11}$\\
        $ 3.60$ & $13 \pm 20$ &$ 9.1 \pm 14.0$ & $3.65 \times 10^{-11}$\\
        $ 4.80$ & $-3 \pm 20$ &$-2.1 \pm 14.0$ & $2.55 \times 10^{-11}$\\
        \hline
    \end{tabular}
    \caption{ Limits on the $n \rightarrow n' \rightarrow n$ regeneration probability per neutron at 95\% CL \cite{Feldman:1997qc} with cadmium absorber for different nominal values of magnetic field settings. Data of 2024 run.}
    \label{tab:counts2024}
\end{table}

The data of the 2021 run were less informative due to the much higher detector background rate 1.5--3.5 cps in the smaller ROI. This higher background rate was caused by unanticipated detector activation during beam intensity calibrations early in the measurement campaign, during which the high intensity beam (although strongly attenuated) was allowed in the detector. 
In the analysis, the background of ROI has an exponential decay time constant of $\sim$3.7 hours, consistent with the $\sim$2\% content of $^{55}$Mn in the stainless steel material of the detector tubes. This background also revealed the sensitivity to the detection of $\gamma$s due to the GP-SANS detector threshold, which was historically set too low at the GP-SANS facility. The background variation in the ROI of the detector during the 40 hour 2021 run is shown in Figure~\ref{fig:background2021}. In the middle of the run, around hour 20, the detector was shifted +20 cm horizontally to a new position, and an attenuated high-intensity beam was impinged on the detector for the short time, thus again activating the detector. Despite a high detector counting rate, it was possible to fit the time-dependent background measured at $B=0$ and subtract it from the measurements with $B\neq 0$ following the same procedure for the background fit and subtraction as for the 2024 data \cite{multinormal}. The corresponding results, normalized by the intensity per neutron in the spectrum and corrected for detection efficiencies, together with the 95\% CL on the regeneration effect, are shown in Table \ref{tab:counts2021}. 
The detector Counts was divided by the general 2021 normalization factor $(2.19 \pm 0.15) \times 10^{12}~\text{h}^{-1}$ to determine 
the signal, representing the probability of effect per average neutron in the beam with definite polarization.

The limits on the probabilities of the regeneration effect at different values of magnetic field, shown in Tables \ref{tab:counts2024} and \ref{tab:counts2021}, are the main experimental results of this paper. We consider these as the result of two independent experiments. 
In both datasets, the uncertainty on the signal for each magnetic field is dominated by the contribution of the uncertainty in the detected counts above backgrounds.

\begin{table}[ht]
    \centering
    \begin{tabular}{|c|c|c|c|}
        \hline
        B-Field (T) & Counts, $h^{-1}$ & Signal ($\times 10^{-12}$) & $95$\% CL \\
        \hline
        $ 2.50$ & $-10.6 \pm 35.0$ & $-4.90 \pm 16.1 $ & $2.67 \times 10^{-11}$\\
        $ 5.00$ & $52.2 \pm 38.8$ &$ 23.9\pm 17.9 $ & $5.88 \times 10^{-11}$\\
        \hline
    \end{tabular}
    \caption{ Limit on the $n \rightarrow n' \rightarrow n$ regeneration probability per neutron at 95\% CL with cadmium absorber for different values of nominal magnetic field settings. Data of the 2021 run. }
    \label{tab:counts2021}
\end{table}

\begin{figure*}
    \centering
    \includegraphics[width=\linewidth]{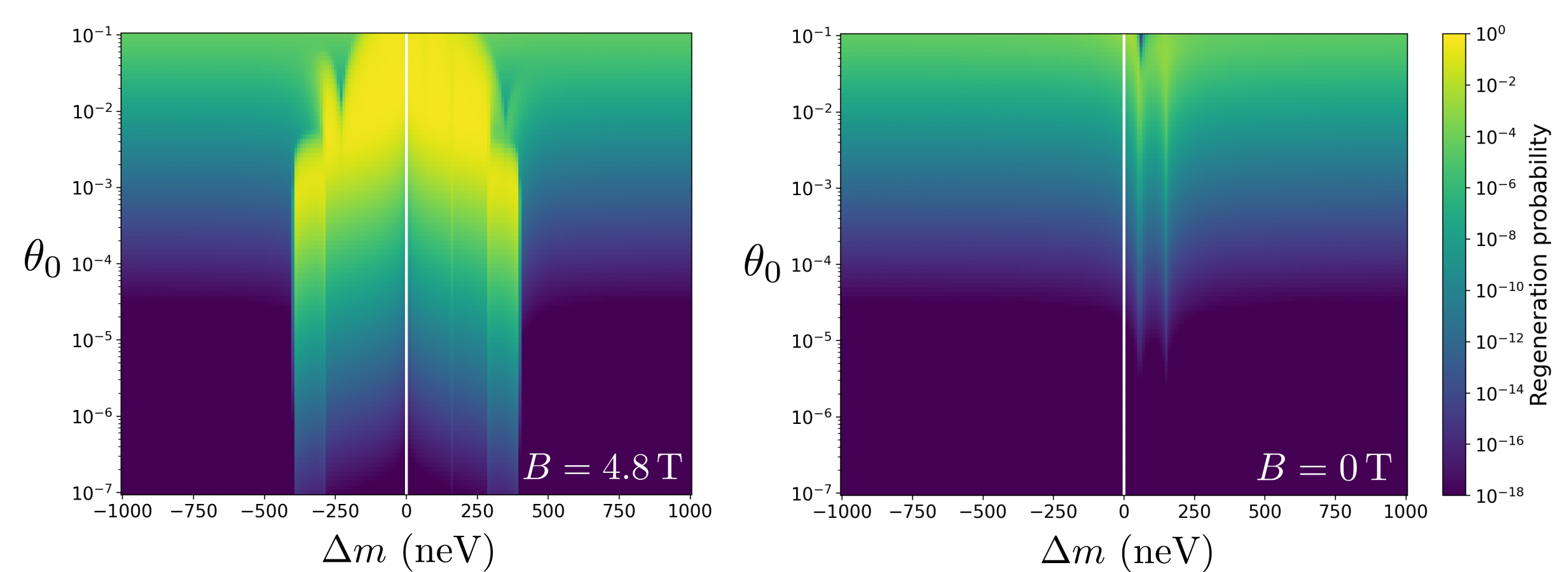}
    \caption{Calculated regeneration probability ($n \rightarrow n' \rightarrow n$) as a function of the $\Delta m$ mass difference and the mixing angle $\theta_0$. Central vertical lines separate two regions from 10 neV to 1000 neV for negative and positive $\Delta m$. Left panel shows the map of probability for nominal \SI{4.8}{\tesla} magnetic field configuration. Right panel is for no magnetic field. See discussion in the text (color online).}
    \label{fig:butterfly}
\end{figure*}

%% file: Section_Text/Results.tex
\section{Discussion of Results}
\label{sec:results}

In two experiments performed in 2021 and 2024, we set upper limits on the regeneration probability for the transformation $n \rightarrow n' \rightarrow n$ within the mixed neutron-mirror neutron model \cite{Berezhiani_2019}, where the mass eigenstates are slightly non-degenerate with the mass splitting $\Delta m$. The theoretical treatment for relating the experimental observables to the parameters of this model, namely $\Delta m$ and the vacuum mixing angle $\theta_0$, was presented in \cite{Gonzalez:2024dba, Kamyshkov:2021kzi}. The principal results of the present study are the 95\% confidence level limits on the regeneration probability, summarized in Tables~\ref{tab:counts2024} and \ref{tab:counts2021} for different values of the magnetic field in the superconducting magnet ~\cite{Magnet} used in the experiment. These probabilities are averaged over the spectrum of cold neutrons in the beam, with wavelengths ranging from 2.5 \AA ~to 20 \AA \cite{Rogers:2024mds}.

To interpret these results in terms of the underlying theory parameters mass difference $\Delta m$ and the mixing angle $\theta_0$, we calculated the average regeneration probability for neutrons on a grid of $(\Delta m, \theta_0)$ combinations. This was achieved using the analytical diagonalization of the two-state non-Hermitian Hamiltonian for piecewise constant fields, as described in Ref. \cite{Kamyshkov:2021kzi}. Starting from a pure neutron initial state, this method produces the survival probability $P_{nn}$ after propagation through the apparatus. Calculations were performed in steps of 30$\mu m$ through the distance of $\pm$ 160 cm from the magnet center where the magnetic field was $>$ 10 Gauss. In each step, the corresponding optical potentials of the materials (air, sapphire, silicon windows, and cadmium absorber) were evaluated, and the magnetic field was interpolated from the known field profile of the magnet. The calculations were averaged over 2000 velocities obtained from McStas simulations of the GP-SANS neutron spectrum, as well as over the two possible neutron spin states relative to the magnetic field in the unpolarized beam. 

We find excellent agreement between this Schrödinger-based calculation \cite{Kamyshkov:2021kzi} and the numerical evolution of the Liouville-von Neumann density matrix used in our previous article \cite{Gonzalez:2024dba}, as expected from the equivalence of the two formalisms. All magnetic field configurations mentioned for the nominal values in Tables~\ref{tab:counts2024} and \ref{tab:counts2021} were considered. As an example, the calculated probability maps for the field configuration \SI{4.8}{\tesla} and for no magnetic field are shown in Figure~\ref{fig:butterfly}. The high-probability regions for neutrons passing the non-uniform magnetic field are due to the non-adiabatic Landau-Zener transition \cite{Landau:1932, Zener:1932ws}. A small effect of probability enhancement is also seen in the absence of magnetic field in the right panel of Figure~\ref{fig:butterfly}. The enhanced regions correspond to the optical potentials of the silicon (54 neV) and sapphire (148 neV) windows in the beam. In addition, the absorption resonance in cadmium (59 neV) as discussed in \cite{Gonzalez:2024dba, Kamyshkov:2021kzi} is seen here for the large mixing angles.

\begin{figure}
    \centering
    \includegraphics[width=\columnwidth]{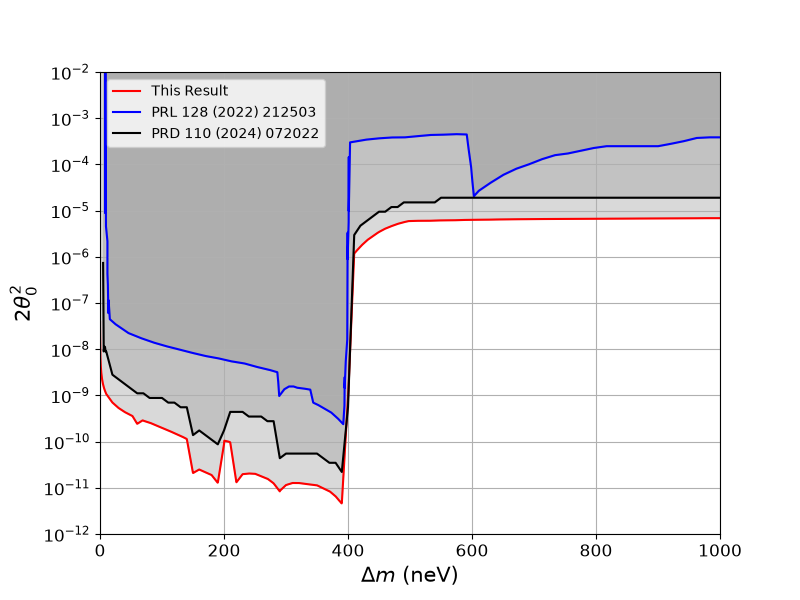}
    \caption{Comparison of the probability limits of two SNS \cite{Broussard:2021eyr,Gonzalez:2024dba} and current HFIR experiments (color online). The gray areas are excluded by the reference experiments.}
    \label{fig:progress}
\end{figure}

The experimental limits in Tables~\ref{tab:counts2024} and \ref{tab:counts2021} were obtained for a given nonzero magnetic field by subtracting the measurement with B=0. Therefore, the calculated probabilities of seeing a signal as a function of parameters $\Delta m$ and $\theta_0$ were also taken from the difference of probability with and without magnetic field for most regions where $\Delta m\lesssim$400\,neV. Because the regeneration transformation becomes adiabatic at $\Delta m>$400\,neV, the probability becomes nearly independent of magnetic field as seen in Figure~\ref{fig:butterfly}, so this approach is not sensitive.  Instead, for large $\Delta m>$400\,neV (and some regions with material resonance), sensitivity limits were taken from the $B=0$ dataset in Table ~\ref{tab:counts2024} comparing the difference of detector counts in the signal ROI and background ROI. Exclusion limits from the different datasets were overlaid to produce the final exclusion limit from the measurements at HFIR.
In Figure \ref{fig:progress} we compare the exclusion limits of our previous experiments \cite{Broussard:2021eyr} and \cite{Gonzalez:2024dba} with the limits obtained in the current experiment (red line) with intense HFIR beam. The vertical coordinate $2{\theta_0}^2$ here represents the probability of a regeneration effect in vacuum. The limits here are shown in the range $\Delta m$ from 10 to 1000 neV. 
For comparison with the existing data from other experiments, we plotted the existing limits in terms of oscillation time 
$\tau$ (s) $=\hbar/|\theta_0 \cdot \Delta{m}|$ for the vacuum transformation $n \rightarrow n'$ for positive and negative $\Delta m$
in the range between $10^{-3}$ and $10^{+3}$ neV from the experiments \cite{Hostert:2022ntu, Ayres:2026lyw,Almazan:2021fvo,Stasser:2020jct,MohanMurthy:2024skb,Ban:2023cja,Brenot:2026ylb}.

%
\begin{figure*}
    \centering
    \includegraphics[trim=100 0 100 0, clip=true, width=\linewidth]{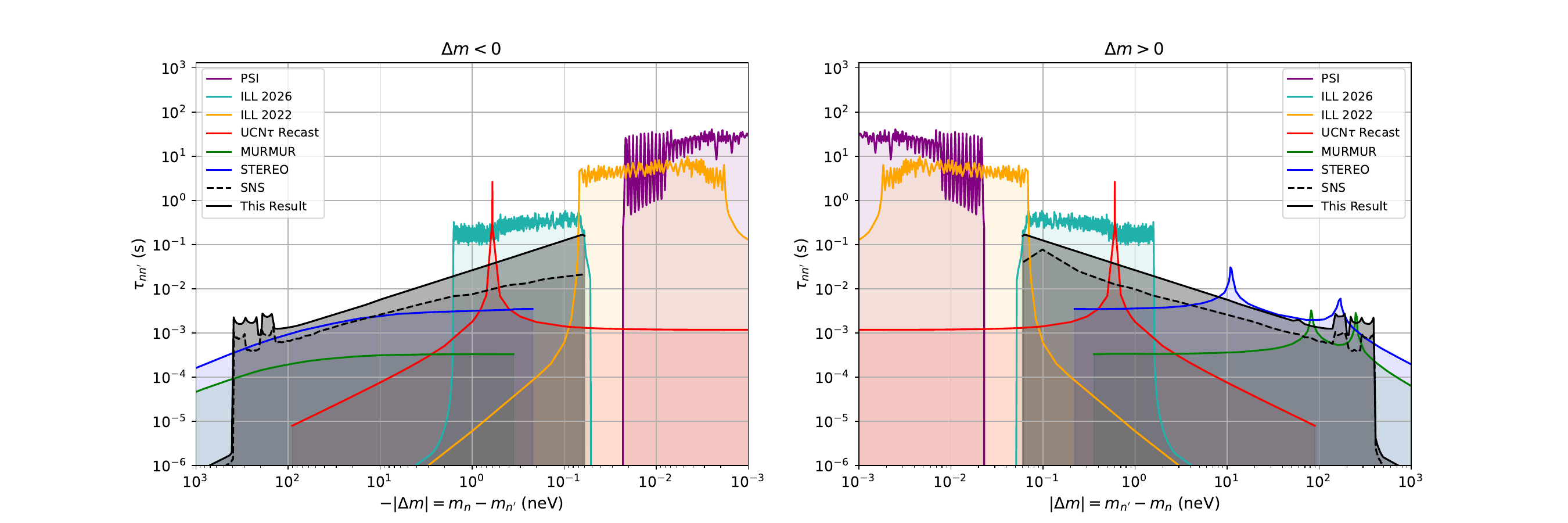}
    \caption{Summary of most recent $\tau_{nn'}$ limits vs positive and negative $\Delta m$ range from $10^{-3}$ neV to $10^{+3}$ neV. Limits reported in this paper (gray area), overlaid with other searches for sterile neutrons, assuming a unified framework as proposed in \cite{Hostert:2022ntu}. Our previous SNS limits are taken from \cite{Gonzalez:2024dba}. The recent 2026 limits marked PSI are results from UCN storage \cite{Ayres:2026lyw, Ziehl:2026kat}. The limit from UCN$\tau$ comes from the non-observation of anomalous losses, as calculated in \cite{Hostert:2022ntu}. UCN beam results marked as ILL 2022 and ILL 2026 use disappearance in the GADGET detector \cite{Ban:2023cja,Brenot:2026ylb}. The STEREO and MURMUR reactor results in terms of $|\Delta m| < 10^{3}$~neV are also presented from \cite{Almazan:2021fvo,Stasser:2020jct}. Limits at lower $\Delta m$ and the UCN $n \rightarrow n'$ limits \cite{MohanMurthy:2024skb} from neutron EDM search apparatus are not shown on this plot.}
    \label{fig:comparison}
\end{figure*}


%% file: Section_Text/conclusions.tex
\section{Conclusions}
\label{sec:conclusions}
This result improves upon the measurements \cite{Broussard:2021eyr} and \cite{Gonzalez:2024dba} performed at ORNL SNS by leveraging the much higher intensity available at HFIR. 
These regeneration type experiments with a super-conducting magnet~\cite{Magnet} explored the model~\cite{Berezhiani_2019} of mixed neutrons and mirror neutrons with transformation $n \rightarrow n' \rightarrow n$, where the masses of the neutron and mirror neutron might be slightly non-degenerate.
In the first publication \cite{Broussard:2021eyr}, we refuted the proposed mechanism~\cite{Berezhiani_2019} of this model to explain the neutron lifetime anomaly. In~\cite{Gonzalez:2024dba} and this paper, we continued exploration of 
a hidden sector, where observation of neutron oscillations can shed light on baryon number violation and the nature of dark matter. Here, we set more stringent limits on the parameters of this model, where the neutron and 
mirror sterile neutron have slightly different masses. An increased sensitivity search for this model will require a cold neutron beam with higher intensity, much longer beam time for measurements, and rather challenging reduction of background in neutron detection.

%% file: Section_Text/acknowledgments.tex
\section{acknowledgments} 
\label{sec:acknowledgments}
The authors thank Prof. Zurab Berezhiani for useful discussions. This research was sponsored by the U.S. Department of Energy (DOE), Office of Science, Office of Nuclear Physics (Contract No. DE-AC05-00OR22725), by the Laboratory Directed Research and Development Program (Project No. 8215) of Oak Ridge National Laboratory, managed by UT-Battelle, LLC, for the U.S. DOE, and in part by the U.S. DOE, Office of Science, Office of Workforce Development for Teachers and Scientists (WDTS) under the Science Undergraduate Laboratory Internship program. The research of the University of Tennessee Knoxville group was partially supported by US DOE Grant No. DE-SC0023149. The University of Kentucky group was partially supported by the US DOE Grant No. DE-SC0014622. 
This research used resources at the High Flux Isotope Reactor (HFIR), a DOE Office of Science User Facility operated by Oak Ridge National Laboratory. The beam time was allocated on the General Purpose Small Angle Neutron Scattering Instrument (GP-SANS) for the proposal numbers IPTS-24916.2 and IPTS-27957.2. This research was supported in part through research cyberinfrastructure resources and services provided by the Partnership for an Advanced Computing Environment (PACE) at the Georgia Institute of Technology, Atlanta, Georgia, U.S. The research at Lund University is supported by Stiftelsen för Strategisk Forskning through the grant RIF21-0057 (Development of a magnetic control beamline at the ESS).